\documentclass[iop,apj,tighten,numberedappendix]{emulateapj}
\usepackage[dvips]{color}
\usepackage{amssymb}
\usepackage{amsmath}
\usepackage{epsfig}
\usepackage{graphics}
\usepackage{natbib}
\usepackage{wrapfig}

\def\deg{$^\circ$}

\newcommand{\be}{\begin{equation}}
\newcommand{\ee}{\end{equation}}

\begin{document}


\title{Influence of the Galactic gravitational field on the positional accuracy of extragalactic sources}

\author{
Tatiana~I.~Larchenkova\altaffilmark{1},
Alexander~A.~Lutovinov\altaffilmark{2,3},
Natalya~S.~Lyskova\altaffilmark{4,2}
}

\altaffiltext{1}{ASC of P.N.Lebedev Physical Institute, Leninskiy prospect 53, Moscow 119991, Russia;}
\altaffiltext{2}{Space Research Institute, Russian Academy of Sciences, Profsoyuznaya 84/32, 117997 Moscow, Russia;}
\altaffiltext{3}{Moscow Institute of Physics and Technology, Moscow region, Dolgoprudnyi, Russia}
\altaffiltext{4}{Max-Planck-Institut f\"ur Astrophysik, Karl-Schwarzschild-Strasse 1, 85741
Garching, Germany}




\begin{abstract}

We investigate the influence of random variations of the Galactic
gravitational field on the apparent celestial positions of extragalactic
sources. The basic statistical characteristics of a stochastic process
(first-order moments, an autocorrelation function and a power spectral
density) are used to describe a light ray deflection in a gravitational field
of randomly moving point masses as a function of the source coordinates. We
map a 2D distribution of the standard deviation of the angular shifts in
positions of distant sources (including reference sources of the
International Celestial Reference Frame) with respect to their true
positions. For different Galactic matter distributions the standard deviation
of the offset angle can reach several tens of $\mu as$ (microarcsecond)
toward the Galactic center, decreasing down to 4-6 $\mu as$ at high galactic
latitudes. The conditional standard deviation (`jitter') of 2.5 $\mu as$ is
reached within 10 years at high galactic latitudes and within a few months
toward the inner part of the Galaxy. The photometric microlensing events are
not expected to be disturbed by astrometric random variations anywhere except
the inner part of the Galaxy as the Einstein--Chvolson times are typically
much shorter than the jittering timescale. While a jitter of a single
reference source can be up to dozens of $\mu as$ over some reasonable
observational time, using a sample of reference sources would reduce the
error in relative astrometry. The obtained results can be used for estimating
the physical upper limits on the time-dependent accuracy of astrometric
measurements.

\end{abstract}


\sloppypar

\section{Introduction}

Trigonometric parallaxes, proper motions and angular sizes of celestial
bodies, which can be determined only by astrometric methods, are the
fundamental quantities in many branches of astronomy and astrophysics. In
particular, proper stellar motions and line-of-sight velocities are nowadays
the main observational data to investigate the structure of the Galaxy, its
gravitational field (the existence of dark matter, the structure of the disk
and of the inner bar, the bulge density), and the evolution of the Galaxy
(the stability of its spiral structure).  The correct handling of astrometric
problems requires a reliable celestial reference system, the implementation
of which is the coordinate system. A standard reference system in astrometry
is the International Celestial Reference Frame (ICRF) constructed by the
position measurements of 212 `defining' extragalactic sources
(http://hpiers.obspm.fr/icrs-pc/icrf/catalogues/icrf.def). These `defining'
sources have at least 20 observations with a total duration of at least two
years. Quasars and distant galaxies are ideal reference sources for defining
the celestial reference system, since their angular motion is $\sim 0.01$ mas
(millisecond of arc) per year \citep{b1}.

In the nearest future, modern technologies will allow us to conduct extremely
accurate radio interferometric observations with an angular resolution of 1
$\mu$as and optical observations with an accuracy of 10 $\mu$as per year.
Analyzing such precise observations, we have to take into account general
relativistic effects associated with the propagation of the electromagnetic
waves in non-stationary gravitational fields. In this regard, it is important
to consider variations in apparent positions (jitter) of  extragalactic
sources due to the propagation of their emission in the non-stationary
gravitational field of both visible stars in the Galactic disk/bulge and
invisible massive objects in the halo. Due to its high importance, this topic
has been actively investigated since the 1990's. The
astrometric microlensing caused by stars in the Galaxy was discussed, for
example, in \citet{b3,b4,b5,b117,b19,b20,b21}, and for binary systems in
\citet{b22,b23,b24,b25}. This is only a short list of the available
literature.

It is important to note that any source belonging to the reference coordinate
system ICRF is also affected by the jitter of its apparent position. In
particular, the improving precision of astrometric observations leads to the
fact that the `jittering' of the reference sources can become visible. It is
a gravitational noise which does not allow us to increase the accuracy of the
implementation of the coordinate system above a certain level. Therefore,
there is a natural limitation on the observational accuracy.

Following this discussion, one can say that the measured coordinates of the
source can be treated as random functions of time. The time variation of the
light deflection from the straight line joining the observer and the source
can be considered as a random or stochastic process. This stochastic process
can be described by such statistical quantities as a mathematical
expectation, variance, and correlation (autocorrelation) function. The
considered characteristics of the stochastic process vary depending on the
direction to the source (the closer the line of sight to the Galactic plane,
the greater the number of randomly passing stars) and the distance from the
source (the further the source, the greater the expected number of
passages).

To describe the gravitational noise, we find the basic characteristics
(first-order moments, the autocorrelation function, the power spectral
density (PSD), and its spectral index) of the light deflection angle in the
gravitational field of moving point masses. For this purpose, we follow the
method proposed by \citet{b2}. Namely, the deflection angle is considered to
be a function of time, i.e. of the impact parameter. In order to calculate
the first-order moments, it is necessary to perform the integration over a
statistical sample of bodies deflecting the light ray. This statistical
sample includes the mass, spatial, and velocity distributions of
light-deflecting bodies. Finally, knowing the autocorrelation function of the
process allows us to calculate the time-dependent characteristics, such as a
conditional variance and a conditional standard deviation. For the small
timescales, the conditional variance characterizes the blur of the data for
the observation series, while for the large time scales it describes secular
variations and tends to the value of the total dispersion. Below, we refer
to it as the dispersion, omitting ''total''.

Note, that this task is quite challenging (as is confirmed by the
above-mentioned papers), and we do not pretend here to provide its final and
exact solution. Our numerical approach to the problem is based on some
simplifying assumptions. The main purposes of this work are the following:
\begin{itemize}
\item (1) To calculate basic statistical characteristics, such as the
standard deviation of the apparent offset in positions of sources, the
autocorrelation function, conditional standard deviations, and the PSD
of a stochastic process. This process determined by random passages of
Galactic objects close to a line of sight during astrometric observations
of distant sources. We perform such calculations for the multicomponent
models of the matter distribution in the Galaxy.
\item (2) To construct the Galactic map of the conditional and total
standard deviations as a visual representation.
\item (3) To estimate how the local non-stationary gravitational
field of the Galaxy affects the accuracy of measured positions (coordinates)
of observed sources including reference sources of the ICRF.
\end{itemize}

In Section~\ref{sec:stat}, we provide an expression for the light ray
deflection angle in the gravitational field of moving massive compact bodies
in the Galaxy as well as expressions for the studied characteristics
of the stochastic process. Section~\ref{sec:results} presents the
calculations and describes the method of calculations. All results are
briefly discussed in Section~\ref{sec:discussion and conclusion}. The mass,
velocity, and spatial distributions of deflecting bodies are given in the
Appendices.

\section{Analytical treatment}
\label{sec:stat}

While propagating through the Galaxy, the light ray from a distant
(extragalactic) source experiences multiple events of changing its trajectory
due to the presence of a large amount of moving compact massive objects such
as stars, stellar remnants, brown dwarfs, etc. The variations of the
gravitational field in time created by the Galactic objects (or in other
words, the fluctuations of the galactic matter density) can be treated as a
stochastic process \citep[e.g.,][]{b6}.

Obviously, for a stochastic process, it is impossible to predict its
instantaneous value. Thus the individual realizations of such a process are
described by random functions with values at any given time being random
variables. As stated in the Introduction, its main characteristics are the
standard deviation, the autocorrelation function, and PSD, the expressions for
which can be found, for example, in \citet{b7}. It is worth mentioning
that we assume the Galaxy to be stationary on large scales, while on small
scales there are some small variations due to motions of stars and other
deflecting compact objects.

Let $\alpha(t, m_{a}, \vec X_{a}, \vec v_{a})$ be a function that describes
realizations of the random process of the light ray deflection in the Galaxy,
where $t$ is time, $\vec X_{a}=(x_{a},y_{a},z_{a}), m_{a}, \vec v_{a}$ are
the coordinates, mass and velocity of the {\it a}th deflecting body,
correspondingly. If the mathematical expectation $\langle \alpha(t) \rangle$
is constant, then the random process is stationary, and the standard deviation
can be written as

\begin{equation}
\displaystyle
\sqrt { \langle \alpha^2 \rangle} = \sqrt {\int dm_{a}d\vec X_{a} d\vec v_{a} f(\vec X_{a},m_{a},\vec v_{a}) \alpha^2(\vec X_{a},m_{a},\vec v_{a})} ,
\label{eq1}
\end{equation}
where the angular brackets denote averaging over the statistical ensemble,
and $f(\vec X_{a},m_{a},\vec v_{a})$ is the probability density function.

The autocorrelation function $\Re (t_i,t_j)$ of the function $\alpha(t,
m_{a}, \vec X_{a}, \vec v_{a})$ is

\begin{equation}
\displaystyle
\Re (t_i,t_j)= \langle \alpha (t_i, m_{a}, \vec X_{a}, \vec v_{a}) \cdot  \alpha (t_j, m_{a}, \vec X_{a}, \vec v_{a}) \rangle, \notag
\end{equation}
where $t_i$ and $t_j$ are different moments in time.

For the stationary random process, the autocorrelation function does not
depend on the time instants $t_i$ and $t_j$, instead, it depends on their
difference $\tau=t_i - t_j$, so

\[
\Re (\tau)= \langle \alpha (t+\tau) \alpha (t) \rangle =
\]
\[
\int dm_{a}d\vec X_{a} d\vec v_{a} f(\vec X_{a},m_{a},\vec v_{a})\times
\]
\be
\alpha(t, \vec X_{a},m_{a},\vec v_{a}) \alpha(t+\tau, \vec X_{a},m_{a},\vec v_{a}),
\label{eq2}
\ee
where $t=t_j$.

The conditional standard deviation can be written as

\begin{equation}
\displaystyle
\sqrt { \langle \alpha^2|\tau \rangle} = \sqrt {\langle \alpha^2 \rangle (1-\bar \Re(\tau))} ,
\label{eq1.1}
\end{equation}
where $\bar \Re(\tau)$ is a normalized autocorrelation function.

Our calculations show that the mathematical expectation is indeed constant
for the large time scales. Thus, the stochastic process, describing
variations of the observed extragalactic source position resulting from the
impact of moving compact Galactic objects, is expected to be stationary. Note
that it is an expected result since  the light ray deflection process is
considered in some spatial cone, in which the mean number of incoming
particles (stars) is approximately equal to the number of outgoing ones. Based
on these arguments, for further calculations, the expressions (\ref{eq1}) and
(\ref{eq2}) are used.

The corresponding PSD can be written as

\begin{equation}
S(\omega)= \frac{1}{2\pi} \int\limits_{-\infty}^{+\infty} \exp {(-i \omega \tau)} \Re (\tau) d\tau
\label{eq0}
\end{equation}
where $\omega$ is the frequency of the Fourier decomposition
of the autocorrelation function.

Assuming that the statistical ensemble of deflecting bodies is defined by
uncorrelated parameters associated with the velocities, masses, and spatial
distributions of those deflecting bodies, the probability density can be
approximated by a product of three statistically independent probability
densities:

\begin{equation}
f(\vec X_{a},m_{a},\vec v_{a})=Af(\vec X_{a})f(m_{a})f(\vec v_{a}),\notag
\end{equation}
where the normalization factor $A$ is defined from

\begin{equation}
\int dm_{a}d\vec X_{a} d\vec v_{a} f(\vec X_{a},m_{a},\vec v_{a})=1. \notag
\end{equation}

We assume here that the integration limits  are known and they set the range
of variations of the statistical ensemble parameters.

Mass, velocity, and spatial distributions of the deflecting bodies are
described in Appendix.

\subsection{Deflection angle}
\label{subsec:defl}

In the limit of small deflection angles, the expression for the light ray
deflection angle in the gravitational field of a gravitating body was
obtained by Einstein. To simplify calculations, we assume that velocities of
deflecting bodies are constant in time ($\vec v_{a}=const$) and a distance
between a photon and a deflecting body at a moment of the closest approach is
much smaller than any other distances characterizing the system. We use the
expression for the deflection of a light ray propagating in the gravitational
field of  arbitrarily moving massive bodies from \cite{b14}:

\begin{equation}
\alpha_{a}^i(t)=\frac{4Gm_{a}}{c^2} \frac{1-\vec {k} \vec v_a/c}{\sqrt {1- v_{a}^2 /c^2}} \frac{P_{j}^i r_{a}^j}{|P_{j}^i r_{a}^j|^2}
\label{eq14}
\end{equation}
where $\vec k$ is the unit vector directed from the emitting source to
the observer, $|r_{a}^j|$ is the distance from the {\it a}th deflecting body to
the observer, $c$ is the speed of light, $G$ is the gravitational constant,
$P_{ij}=\delta_{ij}-k_{i}k_{j}$ is the operator of projection onto the plane
perpendicular to $\vec k$, $\delta_{ij}$ is the Kronecker symbol, $i,j=1, 2, 3$.

\begin{figure}
\begin{center}
\includegraphics[width=\columnwidth,clip]{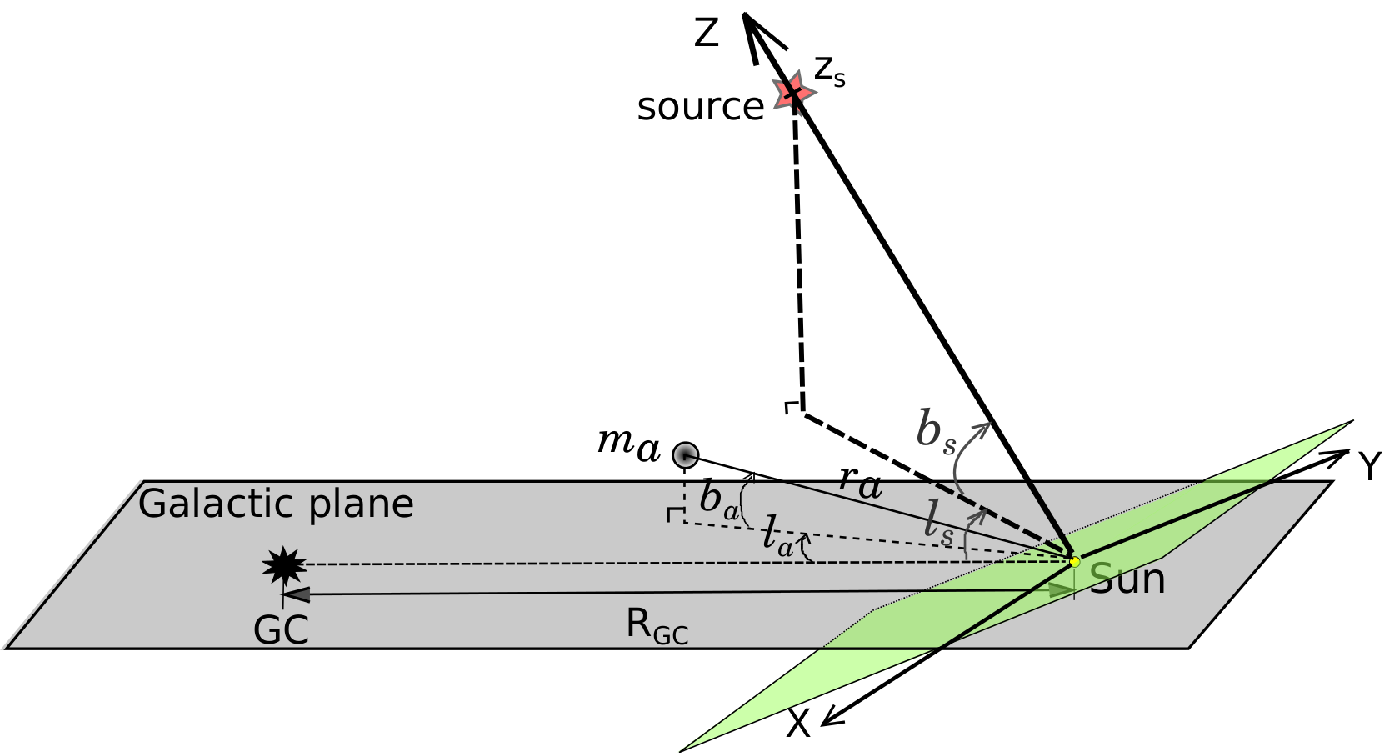}
\caption{Geometry of the problem. $(X,Y,Z)$ is the Cartesian coordinate system
with the $Z$-axis pointing toward the source. $R_{\rm GC}$ and $r_a$ are the
distances from the Sun to the Galactic center and to the {\it a}th
deflecting body, correspondingly. $l_a$ and $b_a$ are the galactic longitude
and latitude of the {\it a}th deflecting body, $l_s$ and $b_s$ are the
source galactic coordinates.
\label{Fig.1}}
\end{center}
\end{figure}

Components of the light deflection by the {\it a}th body in the Cartesian
coordinate system $(x, y, z)$ with the center at the observer position and the Z-axis
directed toward the source (see Fig.\ref{Fig.1}) are defined by

\begin{equation}
\alpha_{a}^{\it x}(t)=\frac{4Gm_{a}}{c^2} \frac{1-\vec {k} \vec v_a/c}{\sqrt {1-{v_{a}^2}/c^2}}\frac{- x_{a}}{x_{a}^2 + y_{a}^2} \left(1+ \frac{z_a}{\sqrt{x_a^2+y_a^2+z_a^2}}\right);\notag
\end{equation}

\begin{equation}
\alpha_{a}^{\it y}(t)=\frac{4Gm_{a}}{c^2}\frac{1-\vec {k}\vec v_{a}/c}{\sqrt {1-{v_{a}^2}/c^2}}\frac{- y_{a}}{x_{a}^2 + y_{a}^2} \left(1+ \frac{z_a}{\sqrt{x_a^2+y_a^2+z_a^2}}\right);
\label{eq15}
\end{equation}

\begin{equation}
\alpha_{a}^{\it z}(t)=0 ,\notag
\end{equation}
and the deflection angle is

\begin{equation}
\alpha_{a}(t) = \sqrt {(\alpha_{a}^{\it x}(t))^2 + (\alpha_{a}^{\it y}(t))^2}
\label{eq16}
\end{equation}

\section{Results of calculations}
\label{sec:results}

\begin{figure*}
\includegraphics[height=\textwidth,angle=-90]{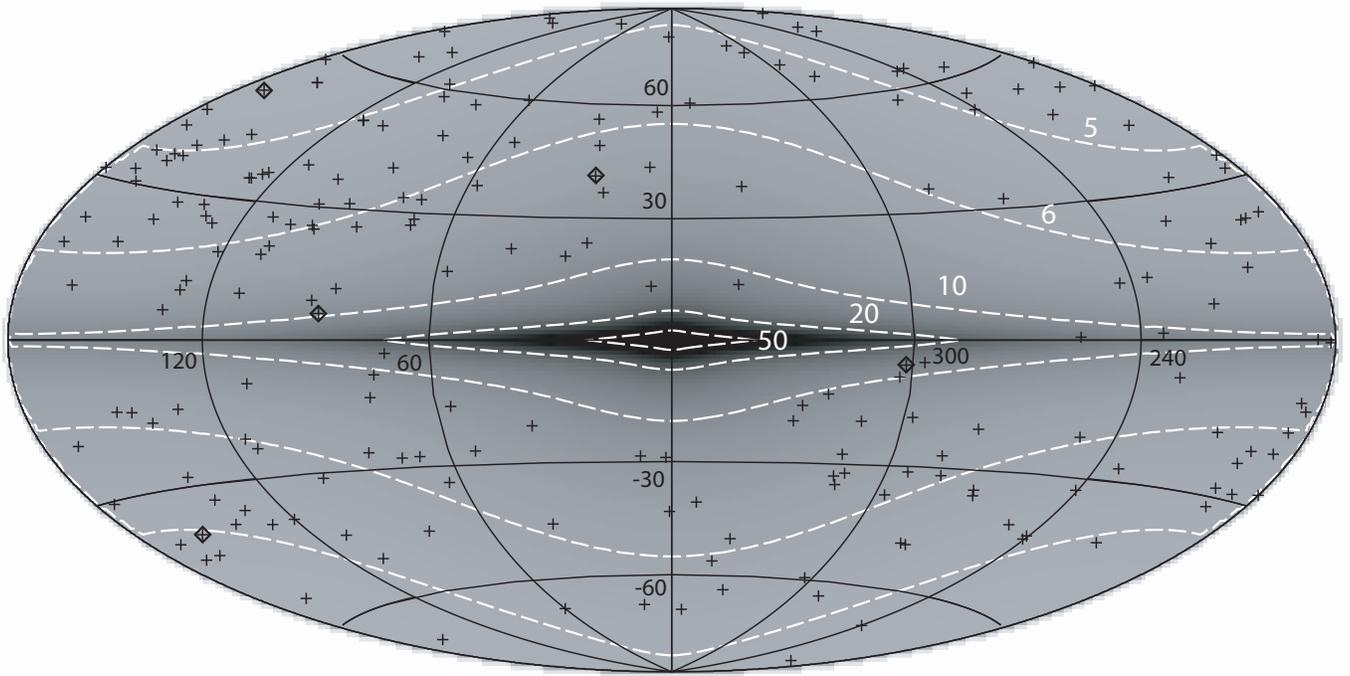}
\caption{Map of the standard deviation of the angle between the measured
and the true source positions on the celestial sphere for the DB model.
Dashed lines show contours of the standard deviation $\sqrt { \langle
\alpha^2 \rangle}$ in $\mu$as. Positions of the ICRF reference sources are
marked as crosses. Several reference sources from Table \ref{tab1} are shown
as diamonds.
\label{Fig.2}}
\end{figure*}
\begin{figure*}
\begin{center}
\includegraphics[height=\textwidth,angle=-90]{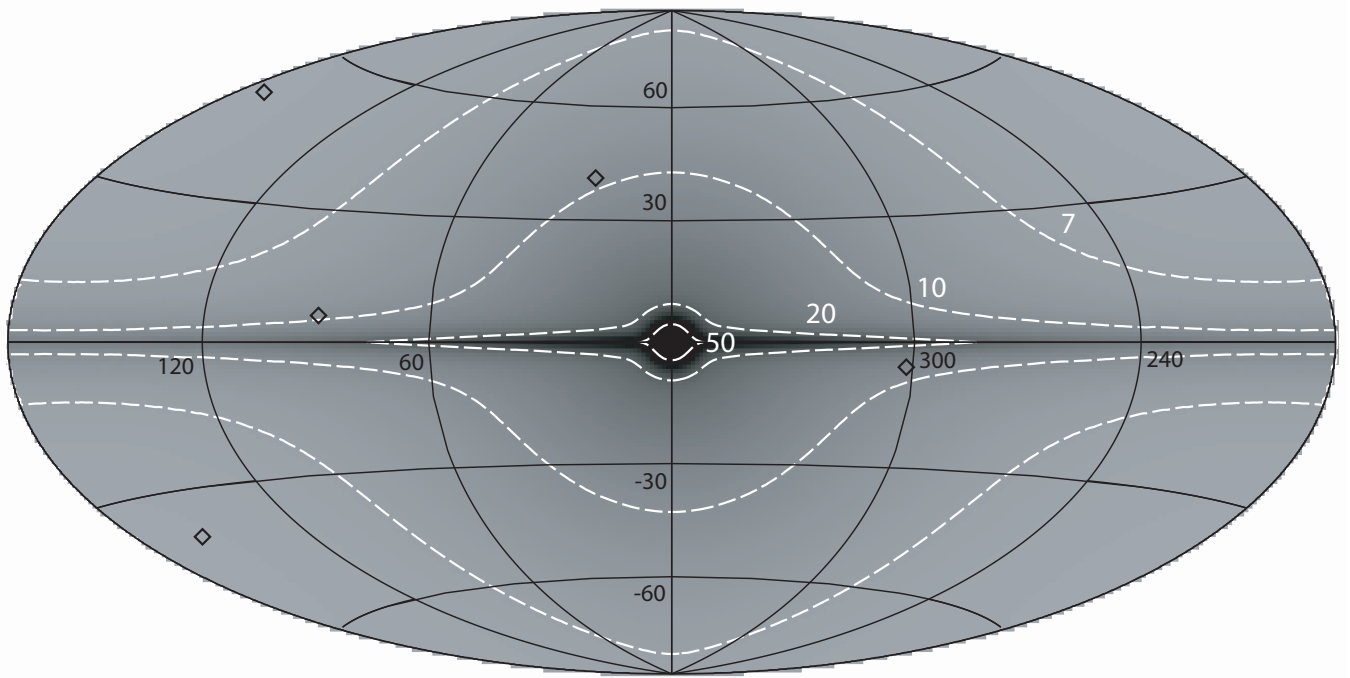}
\caption{Map of the standard deviation of the angle between the measured
and the true source positions on the celestial sphere for the BS model.
Dashed lines show contours of the standard deviation $\sqrt { \langle
\alpha^2 \rangle}$ in $\mu$as. Positions of several reference sources (see
Table \ref{tab1}) from the ICRF catalog are marked as diamonds.
\label{Fig.3} }
\end{center}
\end{figure*}

As mentioned above, the main purposes of this paper are the calculation of the
autocorrelation function and the creation of the Galactic map of the standard
deviation and the conditional standard deviation.

From a mathematical point of view, the calculation of the standard deviation
and the autocorrelation function is reduced to the computation of the
multi-fold integral over the specified limits of integration

\[
\sqrt { \langle \alpha^2 \rangle} =
\]
\be
\sqrt { \sum_{i} \int dm_{a}d\vec X_{a} d\vec v_{a} f_{i}(\vec X_{a})\xi_{i}(m_{a}) f_{i}(\vec v_{a}) \alpha^2(\vec X_{a},m_{a},\vec v_{a})} ,
\ee
\label{eq16.1}

\[
\Re (\tau)=
\]
\[
\sum_{i} \int dm_{a}d\vec X_{a} d\vec v_{a} f_{i}(\vec X_{a})\xi_{i}(m_{a}) f_{i}(\vec v_{a})\times
\]
\be
\alpha(t, \vec X_{a},m_{a},\vec v_{a}) \alpha(t+\tau, \vec X_{a},m_{a},\vec v_{a}),
\label{eq16.2}
\ee
where summation is performed over different components of the Galaxy, and
$f_{i}(\vec X_{a})$, $\xi_{i}(m_{a})$, $f_{i}(\vec v_{a})$ are the spatial,
mass, and velocity distributions of deflecting bodies (see Appendix). Note,
that we use the continuous medium approximation for the complex
multicomponent Galactic structure description.

Let us turn to the discussion of the integration limits in the expressions
(\ref{eq16.1}) and (\ref{eq16.2}). Integration over masses is done within the
limits specified in Appendix A. The integration over the velocity space is
performed up to 500 km s$^{-1}$. When calculating the statistical
characteristics of the stochastic process under consideration, we have taken
into account the influence of all baryonic galactic matter including its
invisible component. The Large and Small Magellanic Clouds have been excluded
from the analysis due to their irregular spatial structure, parametrization
of which is a nontrivial task. For this reason, the integration over $r_a =
\sqrt{x_a^2 + y_a^2 +z_a^2}$ has been done up to 50 kpc.

In this paper, we concentrate on a collective influence of `distant' passages
of stars on the apparent position of the observed extragalactic source, i.e.
only objects with the impact parameters larger than the Einstein--Chwolson
radius\footnote{The Einstein--Chwolson radius is the radius of a ring-like
image of the background source, which appears when the observer, the lens, and
the source are precisely aligned.} have been taken into account. It is well
known that when a star crosses the Einstein--Chwolson cone, the microlensing
effect manifests itself in the form of a source brightness amplification and
a significant displacement of its apparent position (up to milliarcseconds).
Therefore, in our calculations, we have excluded the region with the angular
impact parameters smaller than $\sim3\times10^{-9}$ or $\sim0.6$ mas, what
approximately corresponds to the Einstein--Chwolson radius for a solar mass
star at 20 kpc. The influence of this limitation on the results of
calculations will be discussed below.

\subsection{Galactic map of the standard deviation}
\label{subsec:galmap}

A presence of the supermassive black hole at the Galactic center makes the
structure of this area is fairly complex, which cannot be described by a
single component. Typically, this area is called `the Nuclear Stellar Bulge'
and is believed to be $\sim30 $ pc in size (for more details, see
\citealt{Launhardt.et.al.2002}). This corresponds to an angle of
$\gamma\simeq 0.215^{\circ}$ assuming the solar Galactocentric distance of 8
kpc. The same value of $\gamma$ has been taken as a step in coordinates for
computing the standard deviation of the angle between the measured and the
true positions of the extragalactic source. Taking into account  the fact
that the density distribution of the matter is given by a smooth function
with the characteristic scale of variations significantly exceeding $\gamma$,
the choice of $\gamma$ seems to be justified. For the visualization
purposes, the obtained results have been converted into a so-called AIT
(Aitoff) projection of the celestial sphere in Galactic coordinates and
averaged over $1^{\circ}\times1^{\circ}$.

We have calculated the standard deviation $\sqrt{\left< \alpha^2\right>}$ for
two realistic multicomponent Galaxy models, namely, for the \citet{b12} model
(hereafter, DB model) consisting of a three-component disk, a bulge, and a
halo, and for the `classical' Bahcall-Soneira model (hereafter, BS model)
from \citep{b13.1,b13} which consists of an exponential disk, a bulge, a
spheroid, and a halo (for details see Appendix). Figures \ref{Fig.2} and
\ref{Fig.3} present the maps of the standard deviation for the DB and BS
models, correspondingly. From these maps, it can be seen that the increase of
the standard deviation is expected in the direction to the Galactic center
and the Galactic plane as the matter density increases. In particular, in the
direction to the Galactic center the standard deviation of the deflection
angle reaches several tens of microarcseconds, exceeding 50 $\mu$as in the
central region with the size of $\sim 6^{\circ}\times 2^{\circ}$ for the DB
model, and decreasing down to 4-6 $\mu$as at the high galactic latitudes. The
obtained maps reflect in general the matter distribution in the Galaxy
represented by different models. In particular, the map for the BS model is
characterized by more spherical contour lines than the map for the DB model,
which is caused by the spherical bulge and the stellar spheroid in the BS
model.

\medskip
\begin{table}
\caption{The standard deviation $\sqrt { \langle \alpha^2 \rangle}$ (in
$\mu$as) of the angle between the measured and the true position for several
ICRF reference sources computed for different models of the matter density
distribution in the Galaxy.\label{tab1}}

\vspace{2mm}
\begin{tabular}{c|c|c|c}
\hline

Source Name & $l^{\circ},b^{\circ}$   & $\sqrt { \langle \alpha^2 \rangle}$, DB & $\sqrt { \langle \alpha^2 \rangle}$, BS  \\[1mm]
\hline
J023838.9$+$163659  & 156.7,-39.1 & 4.8 & 5.8 \\[0.5mm]
J095819.6$+$472507  & 170.0, 50.7 & 4.6 & 5.8 \\[0.5mm]
J123946.6$-$684530  & 301.9, -5.9 & 11.4 & 10.9 \\[0.5mm]
J160846.2$+$102907  &  23.0, 40.8 & 6.5 & 9.3  \\[0.5mm]
J203837.0$+$511912  &  88.8,  6.0 & 9.3 & 9.2  \\[0.5mm]
\hline
\end{tabular}
\end{table}
\medskip

Positions of 212 reference extragalactic sources of the ICRF
(http://hpiers.obspm.fr/icrs-pc/icrf/catalogues/icrf.def) are shown also in
Fig.\,\ref{Fig.2}. With increasing an absolute accuracy of space-based
astrometric measurements, it will become necessary to take into account the
discussed jitter of reference sources coordinates caused by the local
non-stationary Galactic gravitational field. Moreover, for about a dozen of
reference sources in the Galactic plane, the standard deviation of the
deflection angle can reach $>10\mu$as. For a visual comparison of the
standard deviations calculated for DB and BS models, we present the standard
deviation values for several reference sources in Table\,\ref{tab1}. The
positions of these sources are marked in Fig.\ref{Fig.2} and \ref{Fig.3} as
diamonds; two sources out of five are located in the Galactic plane (at low
galactic latitudes), and three others at high latitudes. As seen from Table
\ref{tab1} the values of the standard deviations for both models of the
Galaxy are close to each other.

\begin{figure}
\includegraphics[width=\columnwidth,bb=10 100 590 690,clip]{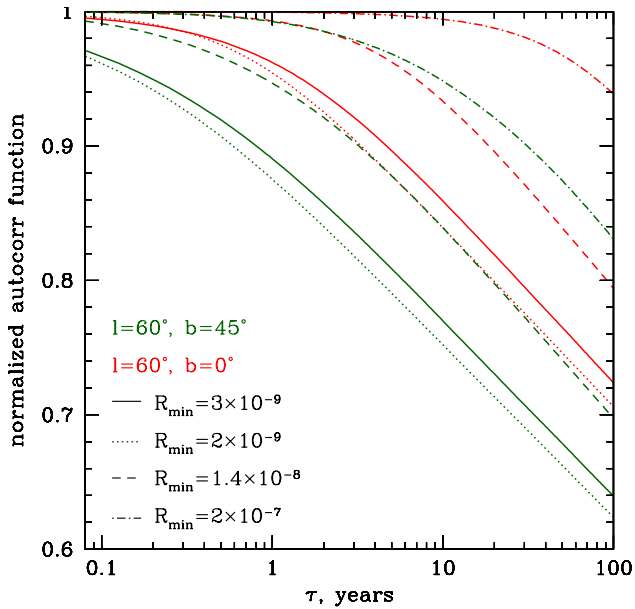}
\caption{Normalized autocorrelation functions calculated for the DB model for two
directions in the sky at different minimal impact parameters.
\label{Fig.4}
}
\end{figure}

As mentioned above, to exclude photometric microlensing events from
consideration, we have introduced the lower limit of integration over
coordinates (i.e., the minimal impact parameter) and have set its value to
$3\times10^{-9}$. We have investigated how obtained results are sensitive to
a particular choice of the this parameter. One order of magnitude variation
of the minimal impact parameter from $2\times10^{-9}$ (this value
approximately corresponds to the Einstein--Chwolson radius $R_{EC}$ for a
solar mass star at a distance of 50 kpc from the observer) to
$1.4\times10^{-8}$ (corresponds to $\simeq R_{EC}$ for a solar mass star at 1
kpc) has introduced only $\simeq$ 10\% variations in the standard deviation.
A further increase of this lower limit up to $2\times10^{-7}$ (corresponds to
$\simeq R_{EC}$ for a solar mass star at 5 pc) lowers the standard deviation
value by about 20\%. Note, that the latter value of the lower limit of
integration clearly exceeds the Einstein--Chwolson radii for the stars in the
Galaxy. Thus, we conclude that the calculated value of the standard deviation
is rather stable against sufficient variations of the minimal impact
parameter.

It is important to note that the maps in Figures \ref{Fig.2} and \ref{Fig.3}
give an order of magnitude estimate of the standard deviation of the
distribution of deflection angles. For practical applications, we still need
to know typical values of the variations, which can be revealed at reasonable
time scales of observations. Corresponding time-dependent statistical
characteristics will be discussed in the next sections.

\subsection{Autocorrelation function, PSD and
conditional standard deviation}
\label{subsec:autocor}

Useful tools for describing stationary processes are an autocorrelation
function and a corresponding PSD, each of which characterizes different
properties of observational data. Therefore, as a first step, we have
computed the autocorrelation function (Eq.\,\ref{eq16.2}) on time scales of
up to 100 years for different directions on the celestial sphere. These
autocorrelation functions can be formally approximated by a superposition of
several components decreasing exponentially with different characteristic
times $T_i$.

\begin{figure}
\includegraphics[width=\columnwidth,bb=20 350 570 690,clip]{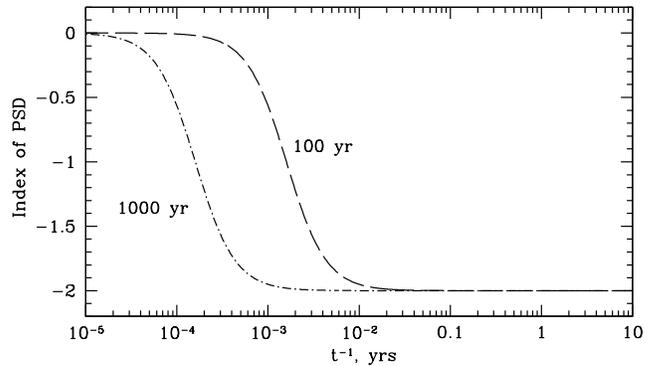}
\caption{Spectral index of PSD calculated for two exponents with the
characteristic time of 100 and 1000 days.\label{Fig.5}
}
\end{figure}

In contrast to the standard deviation, the characteristic decay time of
exponents depends not only on the direction on the celestial sphere but also
on the lower limit of integration over coordinates ($R_{\rm min}$) what can be
seen from Fig.\ref{Fig.4}. Note, that the behavior of autocorrelation
functions for both galactic models (DB and BS) is similar; therefore, in the
following analysis, we are restricted only by the DB model. With increasing
of $R_{min}$, i.e. with increasing the minimal impact parameter, the
characteristic decay time of exponents becomes larger, which seems quite
natural. At the same time, varying the parameters of calculation does not
affect an exponential form of the normalized autocorrelation function.

For the autocorrelation function in a form of a decaying exponent
$e^{-\tau/T}$ the PSD is simply

\be
PSD \propto \frac{T^2}{1 + 4 \pi ^2 T^2 \omega^2},
\label{eq17}
\ee

\medskip
\noindent where $T$ is characteristic time of the exponent, and $\omega =
\tau^{-1}$. From this formula and Fig.\,\ref{Fig.5} it is clearly seen that
on the reasonable observing time-scales (up to one hundred years) the PSD
index equals to -2 for any direction on the celestial sphere and does not
depend on the characteristic time. Thus, if an analysis of the apparent
celestial positions of extragalactic sources, including the reference sources
of the ICRF, reveals a component with a slope of $-2$ in their power spectra,
this component can be explained by the collective influence of stars and
compact objects in the Galaxy.

\begin{figure}
\includegraphics[width=\columnwidth,bb=20 250 580 690,clip]{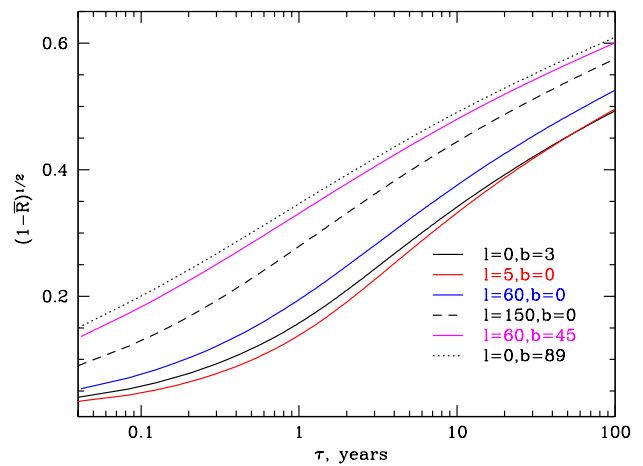}
\caption{Coefficient $\sqrt {1-{\bar \Re}}$ calculated for the DB model for
different directions in the sky with $R_{min}=3\times10^{-9}$. \label{Fig.6}
}
\end{figure}

\begin{figure}
\vbox{
\includegraphics[height=\columnwidth,angle=-90,bb=5 0 415 815,clip]{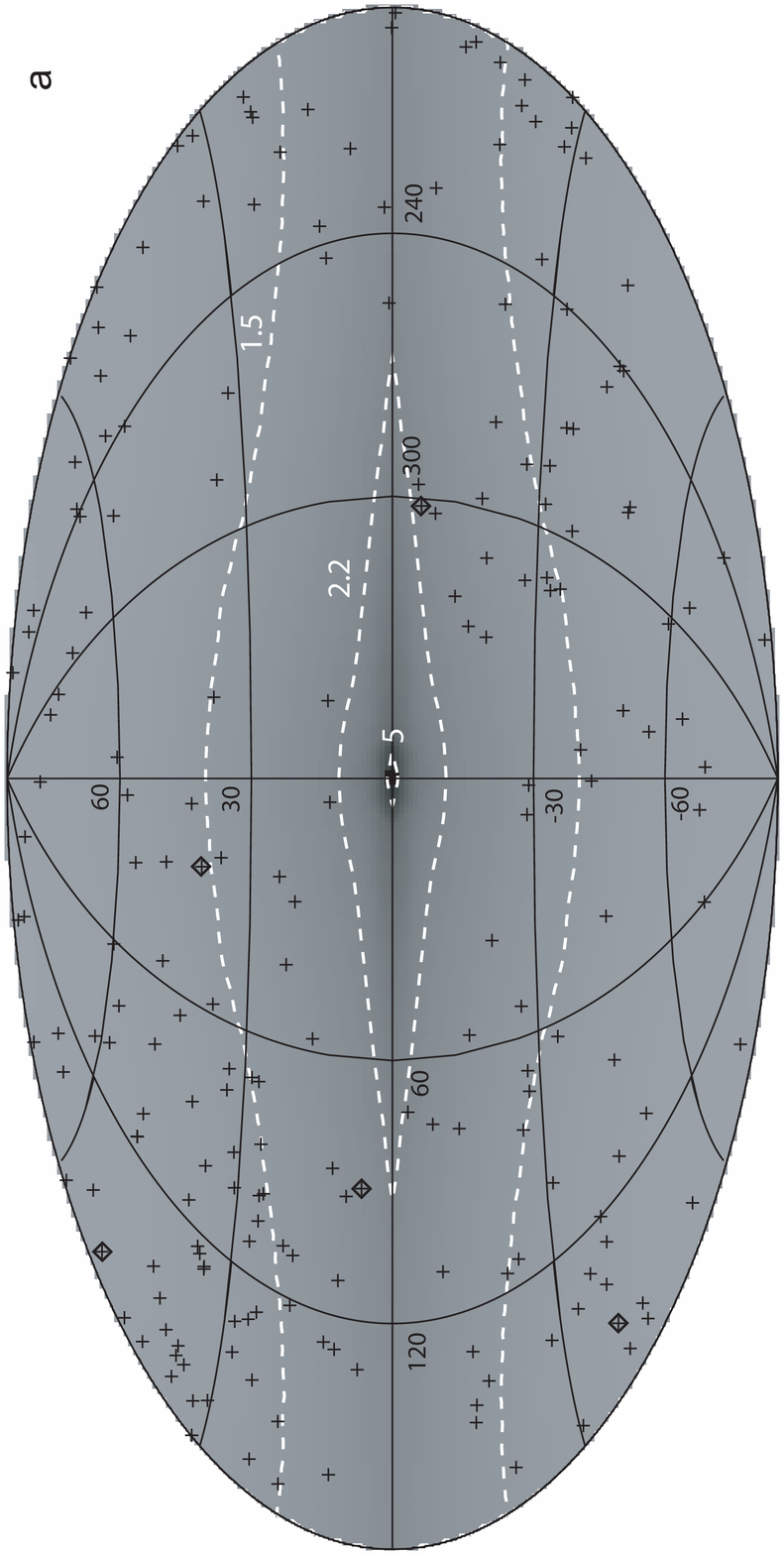}
\includegraphics[height=\columnwidth,angle=-90,bb=5 0 415 815,clip]{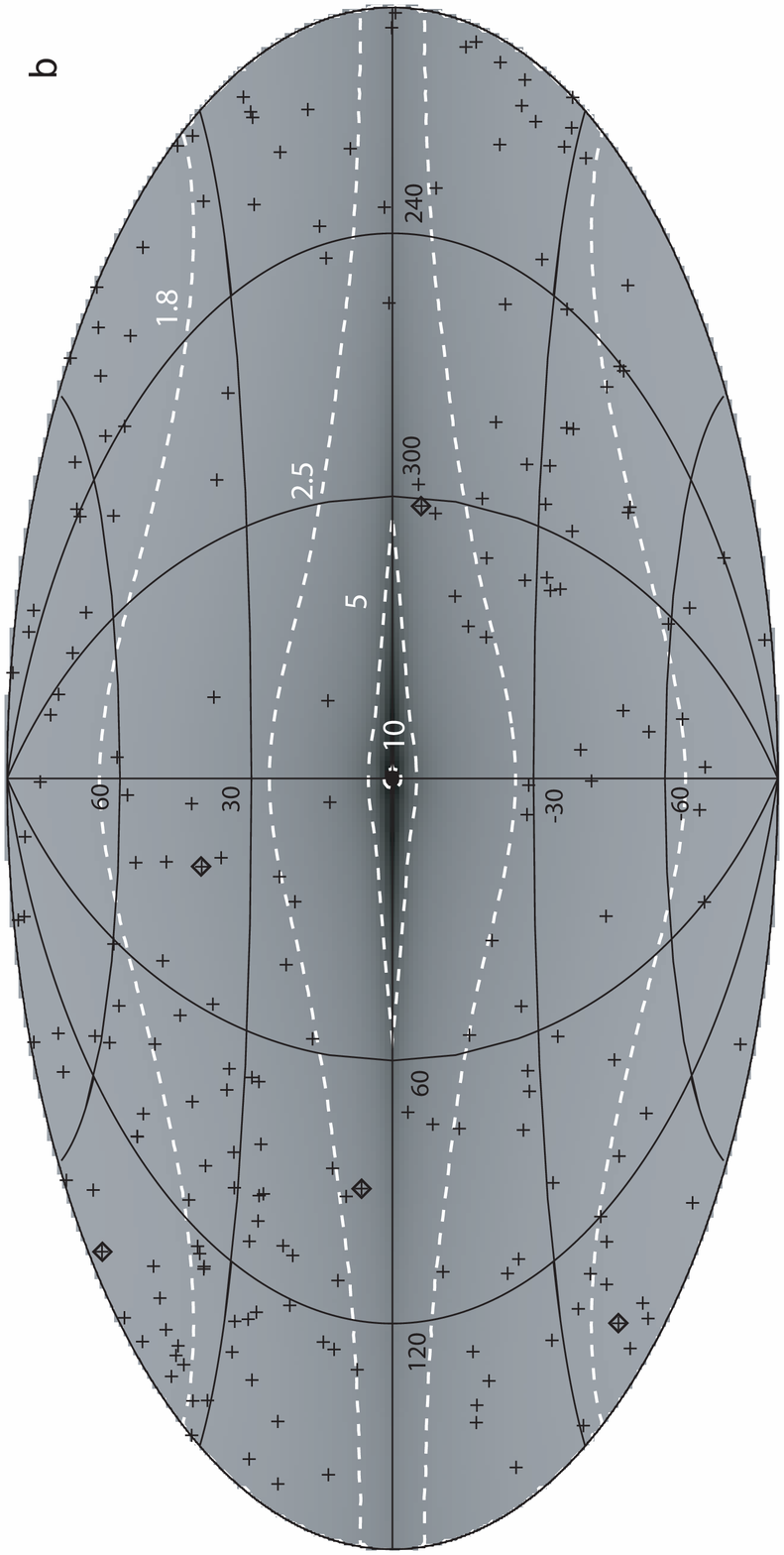}
\includegraphics[height=\columnwidth,angle=-90,bb=170 0 580 815,clip]{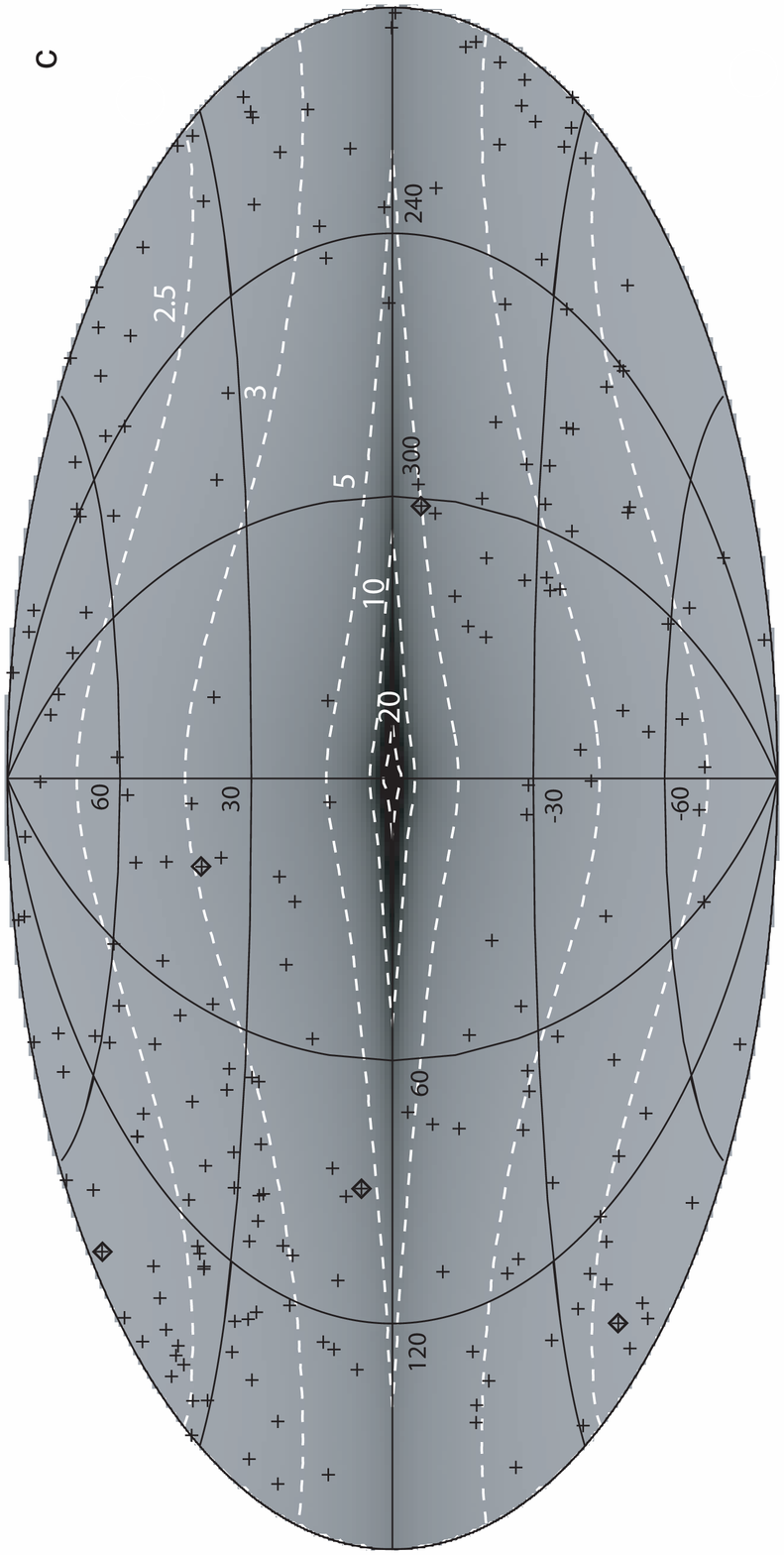}
}
\caption{Map of the conditional standard deviation of the angle between the
measured and the true source positions on the celestial sphere for the DB
model for different observational intervals: 3 months (a), 1 year (b),
10 years (c). Dashed lines show contours of the conditional standard
deviation in $\mu$as. \label{Fig.7}
}
\end{figure}

For the analysis of astrometric observations of staggering precision, it is
important to know an expected value of the jitter of reference sources,
arising due to the local non-stationarity of the Galactic gravitational
field, for a specific time interval. Intuitively, the larger the time
interval between observations, the greater the relative offset of
sources can be. If we know the behavior of the autocorrelation function of the
studied process in time, we can predict the expected value of the jitter
(dispersion) for any time interval. For illustration purposes, the dependence
of the coefficient $\sqrt {1-\bar \Re}$ (Eq.\,\ref{eq1.1}) on time is shown
in Fig.\,\ref{Fig.6}. This coefficient has been calculated for the DB model
for different directions in the sky, and for $R_{\rm min}=3\times10^{-9}$. To
obtain maps of the conditional standard deviation (see Eq.\,\ref{eq1.1}) the
corresponding coefficients were determined for all sky directions for three
time intervals (three months, one year, and ten years) and convolved with the
dispersion map (Fig.\,\ref{Fig.2}). The resulting maps of the conditional
standard deviation are presented in Fig.\,\ref{Fig.7}. From these maps, it is
clearly seen that the jitter may reach a few and a dozen of $\mu as$ in the
direction toward the central parts of the Galaxy at the time scales of 1
and 10 years, respectively, decreasing down to 1\,$\mu as$
at high galactic latitudes.

\section{Discussion and Conclusion}
\label{sec:discussion and conclusion}

In this paper, we have examined the impact of the random variations of the
Galactic gravitational field on the apparent celestial position of
extragalactic sources. The influence of all galactic baryonic matter
including the invisible (non-luminous) components such as brown dwarfs has
been taken into account. The Large and Small Magellanic Clouds were excluded
from consideration due to their complex irregular spatial structure, which is
hard to parameterize. We have obtained the basic statistical characteristics
(mathematical expectation, standard deviation, autocorrelation function, PSD,
and its spectral index) of a stochastic process describing variations of the
extragalactic source positions (coordinates) on the celestial sphere caused
by the deflection of light rays in the gravitational field of randomly moving
point masses. We have mapped the two-dimensional distribution of the standard
deviation of the deflection angle between the measured and the true positions
of distant sources including the reference sources of the ICRF
(Figs.\,\ref{Fig.2}, \ref{Fig.3}). These maps give the order of magnitude of
the expected value of the standard deviation. For these models of the matter
density distribution, it has been shown that in the direction towards the
Galactic center the standard deviation of the deflection angle (jitter) can
reach several tens of $\mu$as, exceeding $50 \mu$as in the very central
region (with the size of $\sim 6^{\circ}\times 2^{\circ}$) and decreasing
down to $4-6 \mu$as at high galactic latitudes.

\begin{figure}
\vbox{
\includegraphics[height=\columnwidth,angle=-90,bb=187 10 595 825,clip]{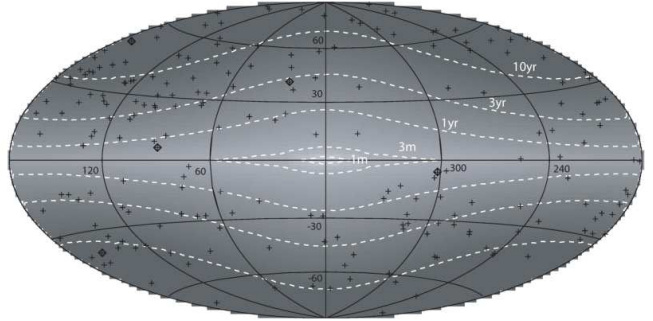}
}
\caption{Map of the jitter time-scale at $ \delta_T = 2.5 \mu as$.
Dashed lines show contours in months and years.
\label{Fig.8}
}
\end{figure}

\medskip
\begin{table}

\caption{The estimated jitter time (in years) at $\delta_T = 2.5 \mu$as for
several ICRF reference sources.}\label{tab2}

\vspace{2mm}
\begin{tabular}{c|c|c}
\hline

Source Name & $l^{\circ},b^{\circ}$  &  $t_{jit,2.5 \mu as}, yrs$ \\[1mm]
\hline
J023838.9$+$163659  & 156.7,-39.1 & $>10$ \\[0.5mm]
J095819.6$+$472507  & 170.0, 50.7 & $>10$ \\[0.5mm]
J123946.6$-$684530  & 301.9, -5.9 & 0.4 \\[0.5mm]
J160846.2$+$102907  &  23.0, 40.8 & 2.6 \\[0.5mm]
J203837.0$+$511912  &  88.8,  6.0 & 0.6 \\[0.5mm]
\hline
\end{tabular}

\end{table}
\medskip

This jitter effect imposes principal limitations on improving the accuracy of
absolute astrometric measurements, making problematic the further improvement
of observation precision due to the emerging `gravitational' noise.
Nevertheless, based on the theoretical estimates of statistical
characteristics of this noise, it is possible to identify it in the analysis
of observational data. The knowledge of the autocorrelation function allow us
to calculate the PSD and the corresponding spectral index
of the `gravitational' noise generated by random passages of Galactic objects
close to the line of sight. It has been shown that the obtained
autocorrelation functions can be formally approximated by a superposition of
several components decaying exponentially with different characteristic
times. For observing time scales up to 100 years the spectral index
of the PSD equals to -2 for any direction on the celestial
sphere and does not depend on (internal) parameters of integration. Thus, if
the analysis of apparent celestial positions of extragalactic sources,
including the reference sources of the ICRF, reveals a component with a slope
of $-2$ in its power spectrum, this component can be explained by the
collective influence of stars and compact objects in the Galaxy and, in
principle, can be extracted.

In practice, it is important to know an expected value of the jitter of
reference sources, arising from the local non-stationarity of the Galactic
gravitational field, for a specific time interval. The autocorrelation
function of the studied process allow us to predict the expected value of the
jitter (the conditional standard deviation) for any time interval. We have
constructed corresponding maps for three observational time intervals (3
months, 1 and 10 years) and showed that the jitter may reach about a few
and a dozen of $\mu as$ in the direction toward the central parts of the
Galaxy at the time scales of 1 and 10 years, respectively, decreasing down
to $\simeq1\,\mu as$ at high galactic latitudes (Fig.\,\ref{Fig.7}).

\begin{figure}
\vbox{
\includegraphics[height=\columnwidth,angle=-90,bb=95 20 505 830,clip]{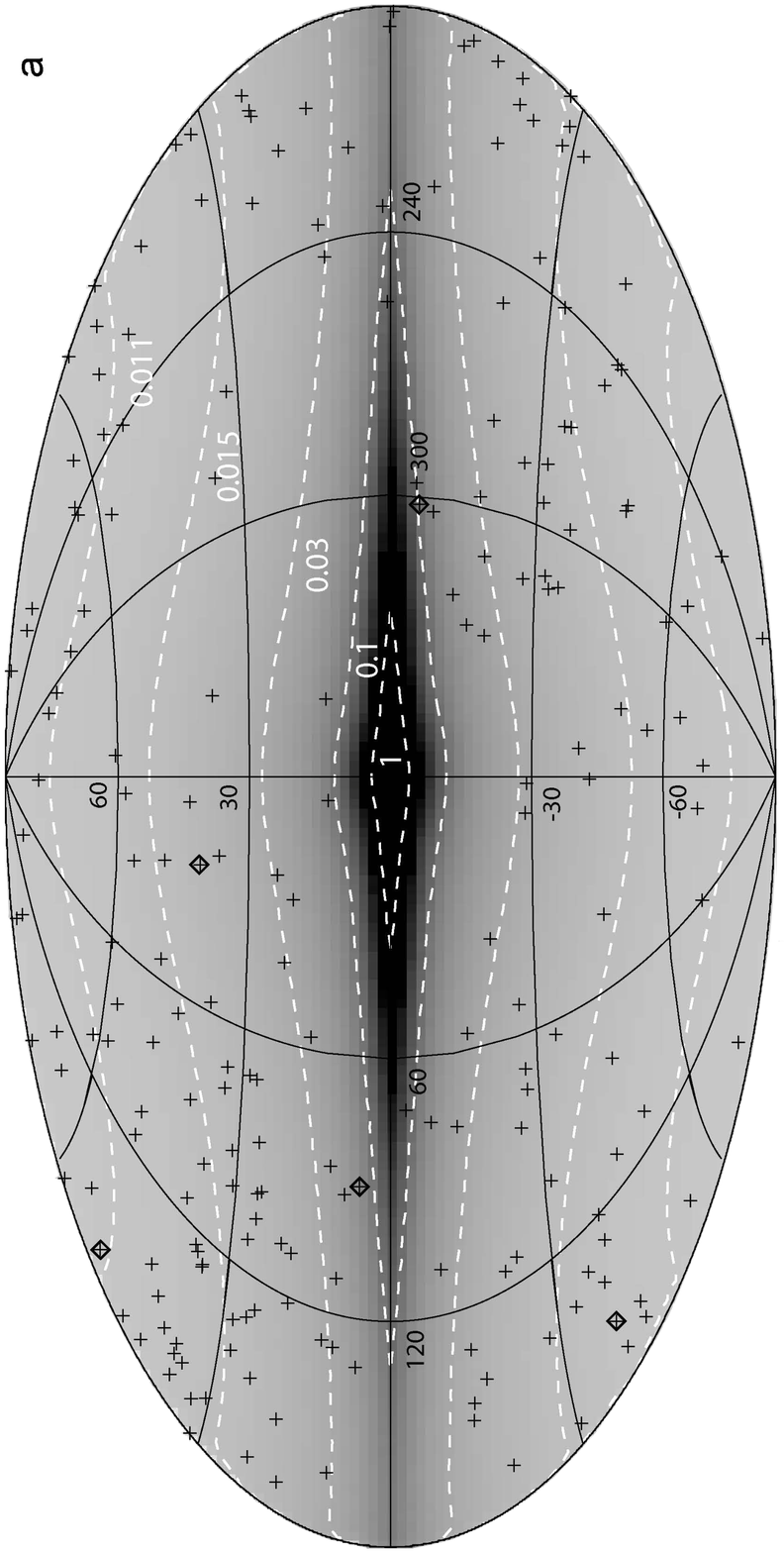}
\includegraphics[height=\columnwidth,angle=-90,bb=180 10 585 820,clip]{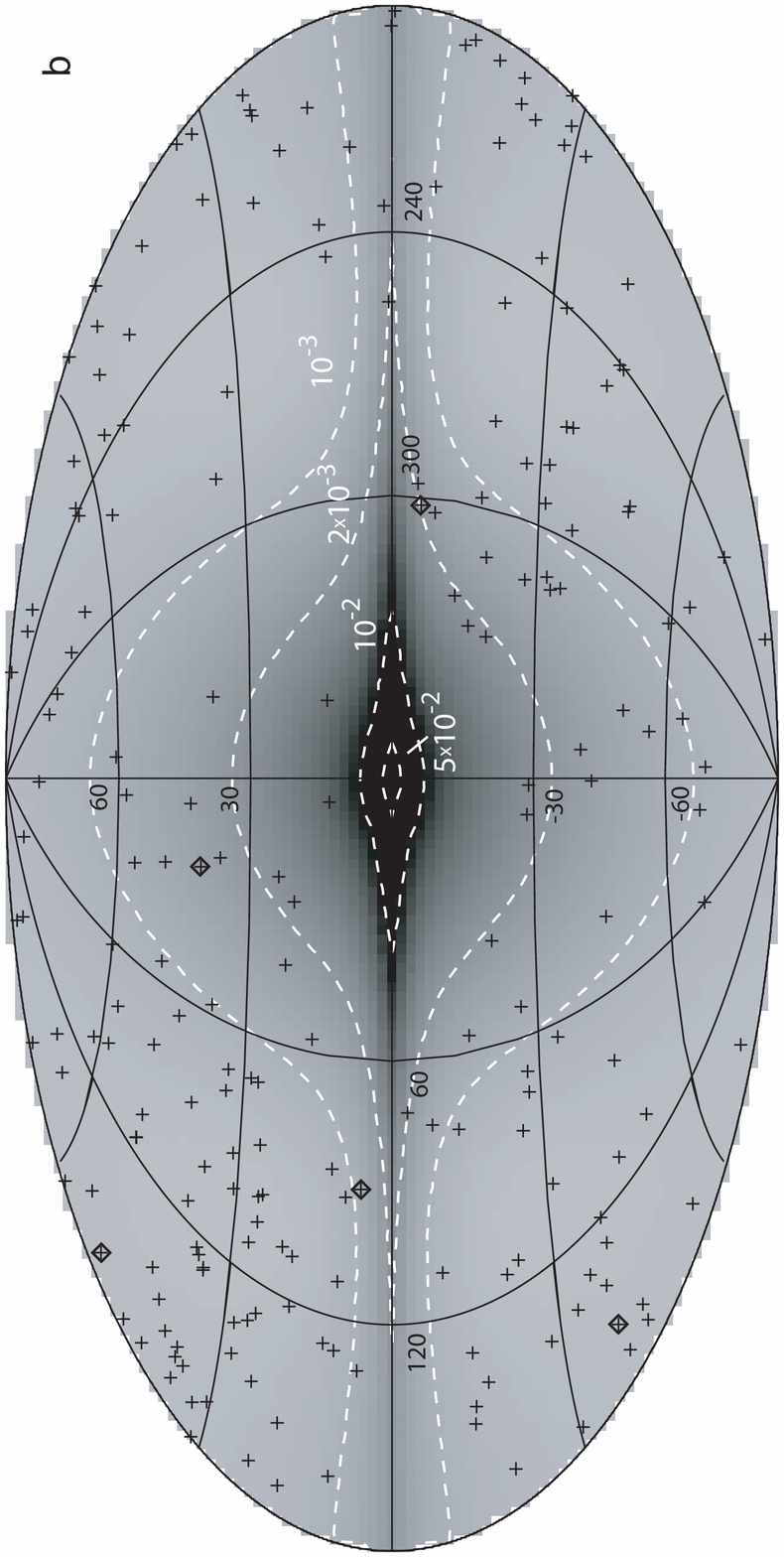}
}
\caption{Maps of the astrometric microlensing optical depth (a) and the
astrometric event rate (b) for the DB model for the threshold $ \delta _T$ =
2.5 $\mu as$. Dashed lines show contours of the astrometric microlensing
optical depth (a) and the astrometric event rate in $yr^{-1}$ (b).
\label{Fig.9}
}
\end{figure}

Additionally, for planning and performing astrometric observations it
would be important to estimate the jitter time in which the local variations
of the Galactic gravitational field will produce an astrometric signature
larger than or equal to a given astrometric precision $\delta_T$. This time
can be estimated from the autocorrelation function by the inverse solution of
Eq.\,\ref{eq1.1}. As an example, we have constructed the map of the expected
jitter times for the astrometric accuracy of $\delta_T = 2.5\,\mu$as
(Fig.\,\ref{Fig.8}). From this map, it is seen that such the jitter can be
produced by random variations of the gravitational field of the Galaxy at
time-scales of 3-10 years at high galactic latitudes ($>30$\deg) and at
time-scales of only several months in the inner part of the Galaxy. The
jitter time estimated for $\delta_T = 2.5\,\mu$as for several reference sources
(the same ones as in Table\,\ref{tab1}) are listed in Table\,\ref{tab2}
for illustrative purposes. The jitter times for other values of the
astrometric accuracy $\delta_T$ can be preliminarily estimated from
Fig.\,\ref{Fig.2} and autocorrelation functions (Fig.\,\ref{Fig.6}).

\begin{figure}
\vbox{
\includegraphics[height=\columnwidth,angle=-90,bb=5 5 415 815,clip]{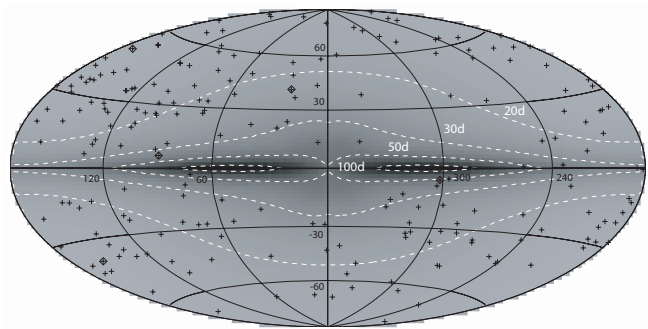}
}
\caption{Map of the Einstein--Chwolson crossing time $\langle t_E \rangle$.
Dashed lines show contours in days.\label{Fig.10}
}
\end{figure}

It is obvious that if the jitter time-scale results are much larger than
the observational time, then the reference sources are `fixed' in space, and
can be used as references for any purpose. If this is not the case, the
associated errors used in reconstructing the astrometry should be taking into
account. At the same time, it is necessary to note that while the jitter of a
single reference source can be up to dozens of $\mu$as over some reasonable
observational time, using a sample of reference sources would greatly reduce
the error in relative astrometry.

We have also calculated the maps of the astrometric microlensing optical
depth and the event rate of extragalactic sources for the detection threshold
$ \delta _T = 2.5 \mu$as. It is important to note that these values for
the microlensing of distant sources are significantly different from ones for
the microlensing of stars in the Galaxy. In particular, Honma \& Kurayama
(2002) demonstrated that the optical depth of astrometric microlensing caused
by disk stars for extragalactic source is larger by an order of magnitude
than that for stars of the Galaxy. Our calculations show that for
extragalactic sources the astrometric optical depth reaches $1$ (i.e., the
probability that a source position shift exceeds a given threshold
$2.5\mu$as is 100\%) and the event rate is $\sim$ 0.05 events yr$^{-1}$ in the
central parts of the Galaxy. Both values decrease by two orders of magnitude
at high galactic latitudes (Fig.\,\ref{Fig.9}). These estimations are
generally agreed with results of calculations performed by other authors
under their assumptions \citep[see, e.g., ][]{b27,b3,2002ApJ...567L.119E}.
The average duration of the astrometric event can be estimated from
corresponding values of the optical depth and the event rate \citep[see,
e.g.,][]{b3}.

Another possibly interesting and important question is whether the jitter has
an impact on the observational properties of sources, in particular, on the
registration of photometric microlensing events. To answer this question, we
have calculated the map of the Einstein--Chwolson crossing time $\langle t_E
\rangle$, which is the time taken for the source to cross the
Einstein--Chwolson radius (Fig.\,\ref{Fig.10}). A comparison of this map with
the time jitter map (Fig.\,\ref{Fig.8}) shows that photometric microlensing
events are not expected to be disturbed by the astrometric random variations
anywhere except the inner part of the Galaxy, as the Einstein--Chvolson times
are typically much shorter than the jittering time-scale for the detection
threshold of $ \delta _T = 2.5 \mu$as.

Summarizing all above, we can conclude that the obtained results can be used
for estimations of the physical upper limits on the  time-dependent accuracy
of astrometric measurements. Therefore, as soon as space-based observations
reach an absolute astrometric accuracy of microseconds of arc, it will be
necessary to take into account the `jitter' of reference source coordinates
caused by the local non-stationary gravitational field of Galaxy.

\bigskip

\section{Acknowledgments}

This work was supported by the Russian Science Foundation (grant 14-22-00271). T.L.
acknowledges a partial support by the program of the Presidium of RAS (P-7) and
the Grant of the President of the Russian Federation for Support of the
Leading Scientific Schools NSh-6595.2016.2. We thank Artur Matveev for his
help with numerical calculations and A.Doroshkevich and D.Novikov for a careful
reading of the manuscript and discussing results. We thank the anonymous referee
for useful comments and suggestions that certainly improved the manuscript.

\bigskip

\appendix

\medskip

\section{Mass distribution of deflecting bodies}
\label{subsec:mass}

The number of stars (and brown dwarfs) per unit mass is the stellar mass
function (MF), which is usually given in linear units (stars $pc^{-3}
M^{-1}_{\odot}$) or in logarithmic units (stars $pc^{-3}
log^{-1}(M_{\odot})$). It is important to note that there is no direct
observational determination of the MF. What is observed is the
individual or integrated light of objects, i.e. the luminosity function or
the surface brightness. Transformation of this observable quantity into the
MF thus relies on theories of stellar evolution (mass--age--luminosity
relations). All stars with main-sequence lifetimes greater than the age of
the Galaxy are still on the main sequence \citep[][]{b7.1}. In that case, the
present-day mass function (PDMF) and the initial mass function (IMF) are
equivalent, in particular, the PDMF and the IMF of the low-mass part of MF
($m\le 1 M_{\odot}$) are equivalent \citep[see, e.g.,][]{b8}.

The PDMFs for stars (including brown dwarfs) of the Galactic disk, bulge,
spheroid, and halo were taken from \citet{b8.1,b8.2,b8} as follows:

\begin{equation}
\label{eq3}
\xi_d(\log m)=
\frac{1}{\log M_{\odot}pc^3}\times\left\lbrace\begin{array}{rl}
0.158 m^0,&0.01 \le m <0.08\\
0.158 \exp \left[ - \frac{(\log \left ( \frac{m}{0.079} \right )^2} {0.9577} \right ],&0.08\le m <1.0\\
0.044 m^{-4.37},&1.0\le m<3.47\\
0.015 m^{-3.53},&3.47\le m<18.2\\
2.5\times10^{-4} m^{-2.11},&18.2\le m<63\\
\end{array}
\right.
\end{equation}

\noindent for the disk;

\begin{equation}
\xi_h (m) = 4\times10^{-3} \left (\frac{m}{0.1 M_{\odot}} \right )^{-1.7} \frac{1}{M_{\odot} pc^3}, 0.01\le m<0.8;
\end{equation}

\noindent for the halo;

\begin{equation}
\xi_{sph}(\log m)=
\frac{1}{\log M_{\odot}pc^3}\times\left\lbrace\begin{array}{rl}
3.6\times10^{-4} \exp \left[ - \frac{(\log \left ( \frac{m}{0.22} \right )^2} {2 (0.33)^2} \right ],& m \le 0.7\\
7.1\times10^{-5} m^{-1.3},&m>0.7.\\
\end{array}
\right.
\end{equation}

\noindent for the spheroid. For the bulge, we used an expression (\ref{eq3}),
but only for $m < 1 M_{\odot}$. Here, $M_{\odot}$ is the mass of the Sun and
$m$ is mass of the star in solar mass units.

The total matter density in the solar vicinity
\begin{equation*}
\rho_{total,solar}=\sum_{i} \int\limits_{m_{\rm min}}^{m_{\rm max}} M \xi_{i}(M)dM , i=d, b, sph, h
\end{equation*}
\noindent is determined in frames of considered models of the Galaxy (see
Appendix C), where $m_{min}$  and $m_{max}$  are the minimal and maximal
masses for each of the galactic components. Note, that the local dynamical
density derived from Hipparcos data is about $0.122 M_{\odot} pc^{-3}$
\citep{b16}.

\section{Velocity distribution of deflecting bodies}
\label{subsec:vel}

For the velocity distribution of deflecting bodies, we adopt the Maxwell
distribution with a cut-off at some $v_e$:

\begin{equation}
f_{i}(v_a) = A_v \frac{\exp \left(-\frac{{v_a}^2}{\sigma_{i} ^2}\right) - \exp \left(-\frac{{v_e}^2}{\sigma_{i} ^2}\right)}{1 - \exp \left(-\frac{{v_e}^2}{\sigma_{i}^2}\right)},
\label{eq4}
\end{equation}
where $A_v$ is a normalization factor and $\sigma_{i}$ is a characteristic
stellar velocity dispersion. In general, its value is different for different
components of the Galaxy; therefore, in following calculations, we used $\sigma
\simeq 100$ km s$^{-1}$ for the halo and spheroid \citep{b9} and $\sigma
\simeq 30$ km s$^{-1}$ for the disk and bulge \citep[see, e.g.,][]{b18,b17}.
The quantity $v_e$ plays a role of the second cosmic speed, i.e. the minimal
speed at which a star is able to escape the local gravitational field of the
Galaxy. According to \citet{b10, b11}  $v_e \simeq 500$ km s$^{-1}$.

\section{Spatial distribution of deflecting bodies}
\label{subsec:spatial}

Because we concentrate here only on the stationary part of the light ray
deflection process caused by the varying in time Galactic gravitational
field. We exclude from consideration the non-stationary part of the effect,
i.e. the influence of individual deflecting bodies crossing the line of
sight, which manifests itself as short-term bursts.

As a plausible model of the spatial distribution of deflecting bodies in the
Galaxy, we use the multicomponent model of the Galaxy from \cite{b12}
(hereafter, DB model) consisting of a three-component disk, a bulge, and a
halo. The distribution of the matter density in the cylindrical coordinates
$(R,z)$ centered at the Galactic center for each of the three components of
the disk is given by

\begin{equation}
\rho_{d}^{DB}(R,z) = \frac{\sum_d}{2z_d} \exp \left( -\frac{R_m}{R}-\frac{R}{R_d}-\frac{|z|}{z_d}\right),
\label{eq5}
\end{equation}
where $R_d$ and $z_d$ are the characteristic length and height of the disk
correspondingly, $\sum_d$  is the central surface density of the disk, and $R_m$
is a parameter describing the decrease in the central surface density of the
interstellar medium. $R_m$ is equal to $0$ for the thin and thick stellar
disks and equals 4 kpc for the disk of the interstellar medium (interstellar
gas).

We exclude the interstellar medium disk from consideration since only the
deflection by compact objects is of interest to us. According to the model 2
from \cite{b12} the total mass of the disk is $4.88 \times 10^{10}
M_{\odot}$, and $R_d=2.4$ kpc, $z_d=180$ pc for the thin disk, and $z_d=1$
kpc for the thick disk.

The bulge and halo density distributions  in the framework of this model are

\begin{equation}
\rho_{b}^{DB}(R,z) = \rho_{0,b} \left(\frac{\sqrt{R^2+z^2/q_{b}^2}
}{R_{0,b}}\right)^{-1.8} \exp\left(-\frac{R^2+z^2/q_{b}^2}{{R_{t,b}^2}}\right)
\label{eq6}
\end{equation}

\begin{equation}
\rho_{h}^{DB}(R,z) = \rho_{0,h} \left(1+\frac{\sqrt{R^2+z^2/q_{h}^2}}
{R_{0,h}}\right)^{-4.207} \left(
\frac{R^2+z^2/q_{h}^2}{{R_{0,h}^2}}\right),
\label{eq7}
\end{equation}
where the central bulge density is $\rho_{0,b}=$0.7561 M$_{\odot}/$pc$^3$, the
characteristic bulge radius is $R_{0,b}=1$ kpc, the truncation radius of the
bulge is $R_{t,b}=1.9$ kpc, the bulge ellipticity is $q_{b}=0.6$, the central halo
density is $\rho_{0,h}=1.263$ M$_{\odot}/$pc$^3$, $R_{0,h}=1.09$ kpc, and the
halo ellipticity is $q_{h}=0.8$.

The matter density in the vicinity of the Sun $\rho _{total,solar}^{DB}$
$\simeq 0.083$ M$_{\odot}/$pc$^3$ for the model under consideration, and the
solar Galactocentric distance is $R_{\rm GC} = 8$ kpc.

The spatial distribution function of compact galactic sources in the DB model
can be written as follows.

\begin{equation}
f_{i}(R,z) = \frac{\rho_{i}^{DB}(R,z)}{\rho _{total,solar}^{DB}}, i=d, b, h
\label{eq8}
\end{equation}

For comparison, we also use another model of the Galaxy density distribution
-- the `classical' Bahcall-Soneira model (hereafter, BS model) from
\citet{b13.1,b13}, which consists of an exponential disk, a bulge, a
spheroid, and a halo. The corresponding density distributions for the
galactic components are given by

\begin{equation}
\rho_{d}^{BS}(R,z) = \rho_{\odot,d}^{BS} \exp \left( \frac{R_{GC}-R}{3.5~kpc}-\frac{|z|}{0.125~kpc}\right),
\label{eq9}
\end{equation}

\begin{equation}
\rho_{b}^{BS}(R,z) = \rho_{0,b}^{BS} \left(\frac{\sqrt{R^2+z^2}}{1~kpc}\right)^{-1.8}
\exp\left[-\left(\frac{\sqrt{R^2+z^2}}{1~kpc}\right) ^3\right]
\label{eq10}
\end{equation}

\begin{equation}
\rho_{sph}^{BS}(R,z) = \rho_{0,sph}^{BS} \frac{\exp \left[-b \left(\frac{\sqrt{R^2+z^2}}{2.8~kpc}\right)^{1/4}\right]}{\left(\frac{\sqrt{R^2+z^2}}{2.8~kpc}\right)^{7/8}}
\label{eq11}
\end{equation}

\begin{equation}
\rho_{h}^{BS}(R,z) = \rho_{\odot,h}^{BS} \frac{a^2+R_{GC}^2}{a^2+R^2+z^2},
\label{eq12}
\end{equation}

\medskip
\noindent
where the central bulge density is $\rho_{0,b}^{BS}=$1.43 M$_{\odot}/$pc$^3$,
the central spheroid density is $\rho_{0,sph}^{BS}= 1/500 \rho^{BS}_{0,d}=\simeq
$0.00079 M$_{\odot}/$pc$^3$, the disk density in the vicinity of the Sun
is $\rho_{\odot,d}^{BS}=0.04$ M$_{\odot}/$pc$^3$, the halo density in the
vicinity of the Sun is $\rho_{\odot,h}^{BS}=0.01$ M$_{\odot}/$pc$^3$, and
$b=7.669$. The core radius of halo $a$ is believed to lie in the range from
$\approx2$ to 8 kpc. We assume in our estimation that $a=2$ kpc.

The spatial distribution function of compact galactic sources in the BS
model is as follows.

\begin{equation}
f_{i}(R,z) = \frac{\rho_{i}^{BS}(R,z)}{\rho _{total,solar}^{BS}}, i=d, b, sph, h
\label{eq13}
\end{equation}

For the BS model $\rho _{total,solar}^{BS}\simeq 0.05 M_{\odot}/pc^3$ \citep{b13.1}.

\vspace{1cm}

\label{lastpage}

\bigskip

\bigskip

\begin{thebibliography}{}

\bibitem[\protect\citeauthoryear{Alcobe \& Cubarsi}{2005}]{b17} Alcobe~S., Cubarsi~R. \ 2005, A\&A, 442, 929
\bibitem[\protect\citeauthoryear{Bahcall \& Soneira}{1980}]{b13.1} Bahcall~J., Soneira~R.~M. \ 1980, ApJS, 44, 73
\bibitem[\protect\citeauthoryear{Bahcall}{1986}]{b13} Bahcall~J. \ 1986, ARA\&A, 24, 577
\bibitem[\protect\citeauthoryear{Carney \& Latham}{1987}]{b10} Carney~B.~W., Latham~D.~W. \ 1987, Proceedings of the IAU Symposium, 39
\bibitem[\protect\citeauthoryear{Chabrier}{2003}]{b8} Chabrier~G. \ 2003, PASP, 115, 763
\bibitem[\protect\citeauthoryear{Chabrier \& Mera}{1997}]{b8.1} Chabrier~G., Mera~D. \ 1997, A\&A , 328, 83
\bibitem[\protect\citeauthoryear{Chandrasekhar}{1943}]{b6} Chandrasekhar~S. \ 1943, Rev.Mod.Phys, 15, 1
\bibitem[\protect\citeauthoryear{Chen}{1998}]{b18} Chen~B. \ 1998, ApJ, 495, L1
\bibitem[\protect\citeauthoryear{Dalal \& Griest}{2001}]{b20} Dalal~N., Griest~K. \ 2001, ApJ, 561, 481
\bibitem[\protect\citeauthoryear{Dehnen \& Binney}{1998}]{b12} Dehnen~W., Binney~J. \ 1998, MNRAS, 294, 429
\bibitem[\protect\citeauthoryear{Dominik}{2006}]{b19} Dominik~M. \ 2006, MNRAS, 367, 669
\bibitem[\protect\citeauthoryear{Dominik}{1998}]{b23} Dominik~M.  \ 1998, A\&A, 333, 893
\bibitem[\protect\citeauthoryear{Dominik \& Sahu}{2000}]{b3} Dominik~M., Sahu~K.~C. \ 2000, ApJ, 534, 213
\bibitem[Evans \& Belokurov(2002)]{2002ApJ...567L.119E} Evans, N.~W., \& Belokurov, V.\ 2002, \apjl, 567, L119
\bibitem[\protect\citeauthoryear{Feissel \& Mignard}{1998}]{b1} Feissel~M., Mignard~F. \ 1998, A\&A, 331, 33
\bibitem[Griest(1991)]{1991ApJ...366..412G} Griest, K.\ 1991, \apj, 366, 412
\bibitem[\protect\citeauthoryear{Han et al.}{1999}]{b22} Han~C., Chun~M., Chang~K.  \ 1999, ApJ, 526, 405
\bibitem[\protect\citeauthoryear{Honma \& Kurayama}{2002}]{b27} Honma~M., Kurayama~T. \ 2002, ApJ, 568, 717
\bibitem[\protect\citeauthoryear{Kopeikin \& Sch\"afer}{1999}]{b14} Kopeikin~S.~M., Sch\"afer~G. \ 1999, PhRvD, 60, 124002
\bibitem[\protect\citeauthoryear{Larchenkova \& Kopeikin}{2006}]{b2} Larchenkova~T.~I., Kopeikin~S.~M. \ 2006, AstL, 32, 18
\bibitem[\protect\citeauthoryear{Launhardt et al.}{2002}]{Launhardt.et.al.2002}  Launhardt~R., Zylka~R., Mezger~P.~G. \ 2002, A\&A, 384, 112
\bibitem[\protect\citeauthoryear{Lee et al.}{2010}]{b21} Lee~C.~H., Seitz~S., Riffeser~A., Bemder~R. \ 2010, MNRAS, 407, 1597L
\bibitem[\protect\citeauthoryear{Miller \& Scalo}{1997}]{b7.1} Miller~G., Scalo~J. \ 1979, ApJS, 41, 513
\bibitem[\protect\citeauthoryear{Nucita et al.}{2016}]{b24} Nucita~A.~A., De Paolis~F., Ingrosso~G., Giordano~M., Manni~L. \ 2016, ApJ, 823, 120
\bibitem[\protect\citeauthoryear{Pugachev}{1960}]{b7} Pugachev~V.~S. \ 1960, The theory of stochastic functions, Moscow
\bibitem[\protect\citeauthoryear{Sajadian}{2015}]{b25} Sajadian~S. \ 2015, AJ, 149, 147
\bibitem[\protect\citeauthoryear{Sazhin}{1996}]{b5} Sazhin~M.~V. \ 1996, AstL, 22, 573
\bibitem[\protect\citeauthoryear{Schneider, Ehlers \& Falco}{1999}]{b15} Schneider~P., Ehlers~J., Falco~E.E. \ 1999, Gravitational Lenses,  Berlin, Springer
\bibitem[\protect\citeauthoryear{Smith et al.}{2007}]{b11} Smith~M.~C.,  Ruchti~G.~R., Helmi~A. et al. \ 2007, MNRAS, 379, 755
\bibitem[\protect\citeauthoryear{Xue et al.}{2008}]{b9} Xue~X.~X., Rix~H.~W., Zhao~G. et al. \ 2008, ApJ, 684, 1143
\bibitem[\protect\citeauthoryear{Yano}{2012}]{b117} Yano~T.\ 2012, ApJ, 757, 189
\bibitem[\protect\citeauthoryear{Zhdanov \& Zhdanova}{1995}]{b4} Zhdanov~V.~I., Zhdanova~V.~V. \ 1995, A\&A, 299, 321
\bibitem[\protect\citeauthoryear{Zoccali et al.} {2000}] {b8.2} Zoccali~M., Cassisi~S., Frogel~J.~A. et al. \ 2000, ApJ, 530, 418
\bibitem[\protect\citeauthoryear{van Leeuwen}{2007}]{b16} van~Leeuwen~F. \ 2007, Hipparcos, the New Reduction of the Raw Data, Astrophysics and Space Science Library, Vol. 350, Berlin: Springer


\end{thebibliography}
\end{document}